\documentclass[a4paper,twocolumn]{article}
\usepackage{graphicx}
\usepackage{tabularx}
\usepackage{array}
\usepackage{rotating}

\newcolumntype{Y}{>{\centering\arraybackslash}X}
\newcommand\RotText[1]{\rotatebox{90}{\parbox{3.5cm}{\centering#1}}}

\title{GenAIOps for GenAI Model-Agility}
\author{Ken Ueno, Makoto Kogo, Hiromi Kawatsu, \\ Yohsuke Uchiumi, and Michiaki Tatsubori\\
IBM Japan
}
\date{}

\begin{document}
\maketitle

\begin{abstract}
AI-agility, with which an organization can be quickly adapted to its business priorities, is desired even for the development and operations of generative AI (GenAI) applications. Especially in this paper, we discuss so-called GenAI Model-agility, which we define as the readiness to be flexibly adapted to base foundation models as diverse as the model providers and versions. First, for handling issues specific to generative AI, we first define a methodology of GenAI application development and operations, as GenAIOps, to identify the problem of application quality degradation caused by changes to the underlying foundation models. We study prompt tuning technologies, which look promising to address this problem, and discuss their effectiveness and limitations through case studies using existing tools.

\vspace{0.5cm}
\end{abstract}
\vspace{0.5cm}

\section{Introduction}

Companies want AI agility so they can gather the necessary resources to build AI capabilities in line with business priorities, and generative AI is no exception. In recent years, the use of generative AI has accelerated among individuals and companies, and the direction of development of foundational models, including large-scale language models (LLMs), is expected to be between general-purpose proprietary models and open models specialized for individual use cases, with these two types of models being used interchangeably. Individual users often use general-purpose models as is, while corporate users are likely to incorporate use-case-specific models into their business applications.

When using multiple models for each function, application development will require tuning and management of prompts for each model, comparative evaluation testing between models, and regression testing when changing models. Development and testing processes and tools are required that enable the agility of generative AI to quickly and efficiently switch models in response to rapid changes in customers, markets, and business. For the development and operation of machine learning models and generative AI platform models, MLOps~\cite{Makinen+:ICSE-WAIN-2021:MLOps} and LLMOps~\cite{Huang+:2024:LLMOps} have been advocated and various tools are available, but there is no established framework or tool for a business application development process that can switch between multiple models in an agile manner.

In this paper, we define a comprehensive framework for an agile generative AI development process in Chapter 2. In Chapter 3, we summarize the issues with model switching that we call "generative AI model agility." In Chapter 4, we survey existing research and tools and discuss solutions to the problem. In Chapter 5, we verify model switching using actual tools and data. Finally, we present solutions to the problem and their limitations.

\section{GenAIOps}

In this section, we define the generative AI development and operation process, which is a prerequisite for discussing AI agility.[14] Although there is no standardized process for machine learning such as CRISP-DM, proposals such as MLOps~\cite{Martinez-Plumed+:IEEEToKDE-2021:CRISM-DM-20years} and CRISP-ML(Q)~\cite{Studer+:2021:CRISP-ML} have been made that follow CRISP-DM[15] but add machine learning-specific steps~\cite{Amershi+:ICSE-SEIP-2019:SE4ML,Breck+:BigData-2017:SE4ML}. Furthermore, in recent years, LLMOps, a process for developing and operating generative AI models and applications, has also been discussed.

\begin{figure*}[t]
    \centering
    \includegraphics[width=\linewidth]{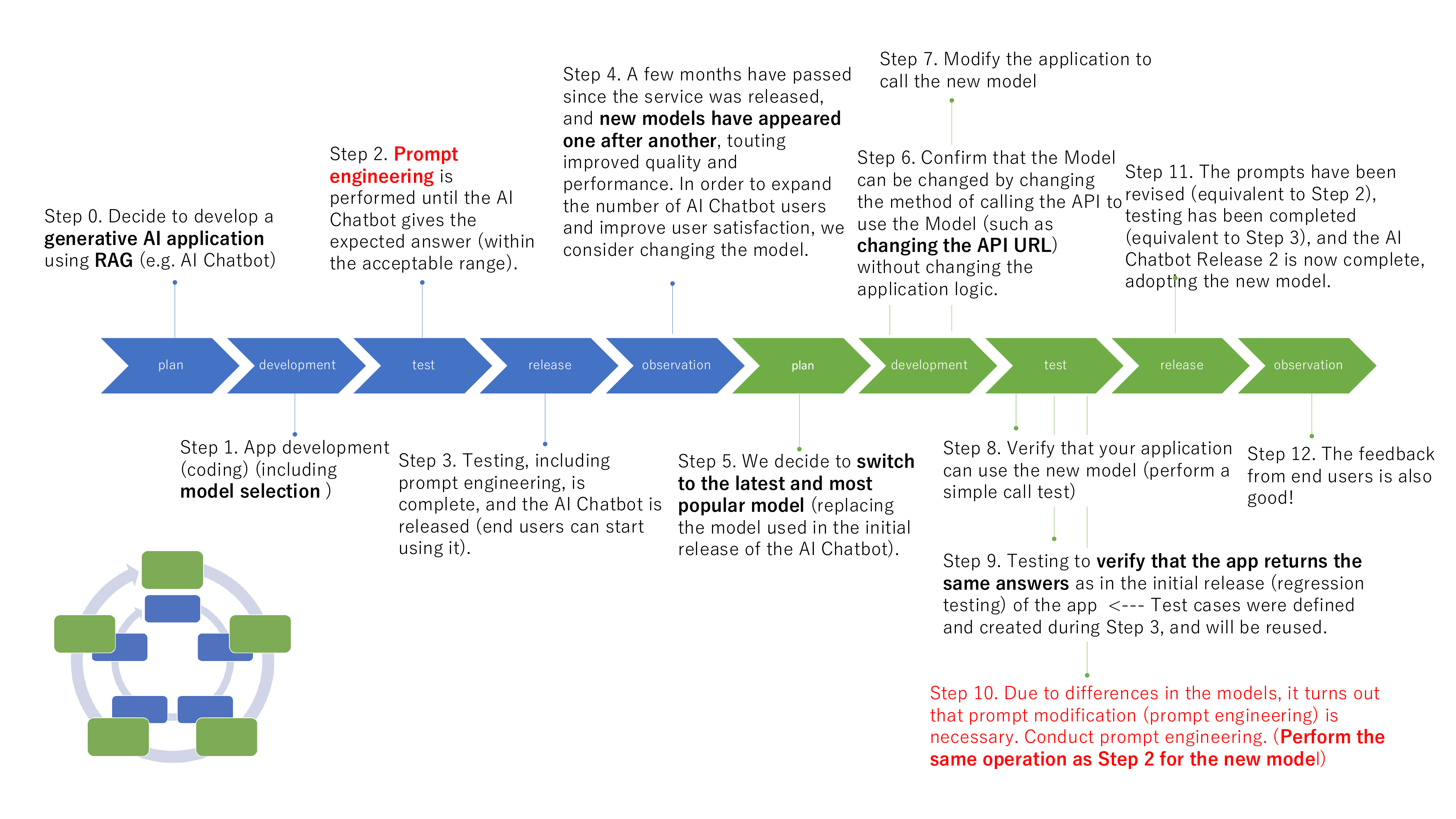}
    \caption{Steps in GenAI application development and operations.}
    \label{fig:GenAIOps}
\end{figure*}

Based on this series of proposals, the authors have defined the flow of LLM application development and operation as shown in Figure 1. Our definition simplifies the definition by integrating multiple steps of the MLOps process so that it is sufficient for discussing the agility of generative AI models, which is our main focus, and then details the aspects specific to generative AI development to highlight problems. Broadly speaking, as shown in Figure 1, we will divide the development lifecycle of an application that includes generative AI into the initial version development phase (blue) and the new version development phase (green).

\subsection{New App Development}

 First, let us look at the flow of development of the initial version. We have broadly categorized the flow of development of a new LLM app (blue part in Figure 1) into five steps.
 \begin{description}
     \item[Plan] The first step (Step 0 in Figure 1) is planning. For example, "Let's develop an app using generative AI!" This is the stage where you decide to "develop a generative AI application using RAG (e.g., AI Chatbot)."
\item[Application Development]
 The next step (Step 1 in Figure 1) is to develop the app (write the code). Here, we will develop the app (write the code), including selecting the LLM model to be used in the application (in this example, an AI Chatbot).
\item[Test]
 The next step after coding (Step 2 in Figure 1) is the testing phase. In LLM app development, prompt engineering is carried out during the testing phase.
 In this example, prompt engineering is repeated until the AI Chatbot gives the expected answer (within an acceptable range). In traditional terms, this is equivalent to repeating tests.
 At the same time, model parameter tuning will also be performed. If this is your first experience with LLM application development, it is difficult to estimate how much time will be required for prompt engineering and model parameter tuning.
\item[Release]
 After testing is complete, the app is released (Step 3 in Figure 1). This is the stage where testing, including prompt engineering and parameter tuning, is completed and the AI Chatbot is released (end users can start using it).
\item[Observation]
 After successfully releasing the first LLM app (AI Chatbot), we enter the observation phase (Step 4 in Figure 1). This is also called the maintenance phase, where we can receive feedback from AI Chatbot users and use it to improve the app.
 This stage is the final step of the first cycle of the life cycle of developing a new LLM app (the final step in blue in Figure 1).
 \end{description}

\subsection{App Maintenance}

Next, the application enters maintenance (the green phase in Figure 1). In this phase, the initial version of the AI Chatbot is released, feedback is received from many users, and then the next version is considered based on that feedback.

\begin{description}
\item[Plan] The AI Chatbot development team begins to consider the next version (Step 5 in Figure 1). This includes discussions on changing to a new model.
 For example, several months after the service release, new models appear one after another, touting improved performance (improved accuracy), and in order to expand the number of AI Chatbot users and increase user satisfaction, they begin to consider changing the model. This is the story of how the AI Chatbot development team decides to change to the latest and most popular model (replace the model used in the initial release of AI Chatbot with the new model).
\item[Development]
 The AI Chatbot development team will identify areas that need to be modified in the app and confirm that the model can be modified without changing the application logic, by simply changing the way the API is called to use the model (e.g., changing the API URL) (Steps 6 and 7 in Figure 1).
 In this example, the story is about improving the app by replacing the model without making major changes to the application logic.
\item[Test]
 To use the newly adopted model, a simple call test is performed to confirm that the model can be replaced simply by changing the method of calling the API (e.g., changing the API URL) (Step 8 in Figure 1).
 After that, we conduct a test (regression test) to confirm that the response returned is the same as that of the initial release of the app. The test cases are defined and created when developing a new app, and are reused (Step 9 in Figure 1).
 The important point here is to implement prompt engineering. Due to differences in the models (the model has been replaced), it becomes necessary to modify the prompts (prompt engineering). In other words, prompt engineering that is appropriate for the new model must be implemented.
 In reality, it is difficult to predict how much work this will entail. It is expected that prompt engineering work equivalent to that required for new development will be carried out for the new model.
\item[Release]
 Once the prompts have been revised (prompt engineering work equivalent to that required for new development), and testing is completed (work equivalent to that required for new development), the AI Chatbot Release 2 adopting the new model is complete (Step 11 in Figure 1).
\item[Observation]
 After the release of the version adopting the new model, feedback from end users was positive and we entered the operation phase (Step 12 in Figure 1).
\end{description}

\subsection{Regression Test}

In this paper, we consider the automation of prompt engineering to reduce the need for prompt modification when changing models used in LLM applications. In order to optimize prompt modification, some kind of training data is required, but in this paper we assume that the contents of regression testing will be used and discuss what kind of testing needs to be performed.

Check whether the application behaves the same (behaves the same from the user's perspective and returns the same answer) before and after the model is changed. This corresponds to the so-called "regression test".   How should regression test cases for LLM applications be created and prepared appropriately?

In web application development, particularly in user interface testing, there are test automation frameworks that emulate web browsers and test whether the same user experience is being provided from the perspective of the website user as before (Selenium is a representative tool used in many development sites), making it easy to detect differences in the appearance of a website depending on the version.

However, in LLM, even if the same prompt is given, the answer is never exactly the same as the previous time, so is traditional regression testing using the tools mentioned above appropriate?

\subsection{Auto prompt-engineering}

Soft prompt tuning technology has been put to practical use to automate prompt engineering. Prompt tuning, as adopted in Watsonx Tuning Studio, etc., performs tuning by adjusting the prompt vectors that are tokenized and vectorized from user input.

Meanwhile, various methods known as automatic prompt engineering~\cite{Zhou+:ICLR-23:autoPE} have been proposed in recent years. Automatic prompt engineering is an optimization technique that aims to have the LLM itself generate prompts that were previously created by humans. For example, the following papers and services have been published:
\begin{itemize}
    \item Automatic Prompt Engineer~\cite{Zhou+:ICLR-23:autoPE}
    \item OPRO~\cite{Yang+:2024:OPRO}
    \item EvoPrompt~\cite{Guo+:2024:EvoPrompt}
    \item ProTeGi~\cite{Pryzant:2023:ProTeGi}
    \item SAMMO~\cite{Schnabel+Neville:2024:SAMMO}
\end{itemize}

Many of these methods have been proposed for the purpose of publishing academic papers, and have only been verified for specific models. In addition, many of them are not sufficiently maintained as reusable libraries, and only a limited number of them are available for continuous use. Furthermore, there remains the challenge that advanced machine learning expertise is still required to effectively optimize these methods. 

\section{Evaluation Framework}

We developed an evaluation framework for the LLM application, which consists of the following three items:
\begin{description}
\item[Functionality] - We evaluate the accuracy of the LLM by measuring whether the output is performed with the accuracy expected by the developer and how well the AI service executes the task. Here, we also check whether the output is in line with the developer's intention, even for unexpected input.
\item[Safety and trustworthiness] - Evaluate whether there is any input or output that could identify an individual, or any output that could potentially harm a person. Also evaluate whether the AI complies with various guidelines.
\item[Fairness]  - We evaluate whether the algorithms and data are unbiased, are socially acceptable from the perspective of diversity, and produce fair output for everyone. 
\end{description}

\newcommand{\Summary}[1]{\parbox{5cm}{\vspace{0.2\baselineskip}#1\vspace{0.5\baselineskip}}}
\begin{table*}[ht]
\centering
\caption{Available evaluation metrics}
\label{tab:evaluation-metrics}
\vspace{5mm}
\scriptsize
\begin{tabular}{|l|l|c|c|c|c|}
\hline
Evaluation Metric & Summary & \multicolumn{4}{c|}{Applicable Tasks} \\ \cline{3-6}
& & \RotText{Summarization} & \RotText{Content Generation} & \RotText{Question Answering} & \RotText{Entity Extraction} \\ \hline\hline
ROUGE & \Summary{Indicates the degree to which a summary or generated text\\ reproduces the original data; recall.} & ◯ & ◯ & ◯ & ◯ \\ \hline
SARI & \Summary{Summarization accuracy; indicates the degree of agreement with the original text; agreement rate.} & ◯ & & &  \\ \hline
METEOR & \Summary{Average of recall and precision.} & & ◯ & ◯ &  \\ \hline
Text Quality & \Summary{Measures F1 score, precision, and recall against model\\ predictions and ground truth data.} & ◯ & ◯ & & \\ \hline
BLEU & \Summary{Indicates the degree of similarity between the generated text\\ and the evaluation target.} & ◯ & ◯ & ◯ &  \\ \hline
Sentence Similarity & \Summary{Converts input text to a vector and calculates its similarity.} & ◯ & & &  \\ \hline
Readability & \Summary{Grades the readability of the generated text on a 7-point scale..} & ◯ & ◯ & &  \\ \hline
Exact Match & \Summary{Indicates the degree of matching to the reference.} & & & ◯ & ◯ \\ \hline
Multi-Label/Class Metrics & \Summary{Measures the performance of a multi-label/class prediction model.} & & & & ◯ \\ \hline
\end{tabular}
\end{table*}

To evaluate the output accuracy of the LLM application, different metrics from watsonx.governance are used depending on the task. For example, in the question answering task, metrics that evaluate recall, similarity, and agreement, such as ROUGE, BLEU, and Exact Match, are used. Table \ref{tab:evaluation-metrics} summarizes the evaluation metrics that can be used for each LLM task.

To assess whether both the input and output data contain personal information, we use the PII (Personally Identifiable Information) index from watsonx.governance. In addition, to assess whether both the input and output data contain violent or inappropriate language, we use the HAP (Hate, Abuse, and Profanity) index.
In addition, assessment templates on watsonx.governance will be used to evaluate compliance with various guidelines. 

Possible causes of bias include bias in the data or samples, and learning bias. To create an AI that produces fair output, various tools can be used in the pre-processing, learning, and post-processing stages of building an LLM. Since this paper does not focus on tuning the LLM itself, it uses a post-processing tool that can evaluate unfairness between segments for the output of an existing AI model. Specifically, we use AI Fairness 360's Reject Option Classification~\cite{Kamiran+:KMIS-11:RejectOptionClassification}, Equalized Odds Postprocessing~\cite{Hardt+:NIPS-16:EqualizedOP}, and Calibrated Equalized Odds Postprocessing~\cite{Pleiss+:NIPS-17:CalibratedEqualizedOP} as evaluation indicators.

\section{Justification of Various Optimization Methods}

Among the various optimization methods that can be applied to LLM, we have actually verified representative methods. In this chapter, we describe the verification results.

\subsection{Prompt Tuning}

Tuning Studio (IBM watsonx) allows for fine tuning of existing models using additional training data, prompt tuning, and other tuning methods. Here, we describe the results of prompt tu[5]ning using the dialogue summary dataset DialogSum.

\begin{table}[hbt]
\caption{Parameters for prompt tuning.}
\label{tab:parameters}
\vspace{\baselineskip}
\begin{tabular}{l|l}
%\hline
Parameter & Value \\ \hline\hline
Task & Generation \\ %\hline
Gradient accumulation steps & 16 \\ %\hline
Batch Size & 16 \\ %\hline
Initialization method & random \\ %\hline
Initialization text & (blank) \\ %\hline
Learning rate & 0.3 \\ %\hline
Max input tokens & 256 \\ %\hline
Max output tokens & 128 \\ %\hline
Number of epochs & 20 \\ %\hline
Number of virtual tokens & 100 \\ %\hline
\end{tabular}
\end{table}

The model was ibm/granite-13b-chat-v2, and two types of training data, 250 and 9950 cases, were prepared, and the changes in optimization step and loss were compared. The main parameters used in the validation are shown in Table~\ref{tab:parameters}.

\begin{figure}[t]
    \centering
    \includegraphics[width=1\linewidth]{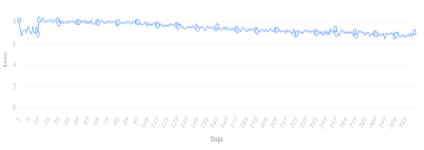}
    \caption{Prompt tuning results (size = 250).}
    \label{fig:tuningresult-250}
\end{figure}

\begin{figure}[t]
    \centering
    \includegraphics[width=1\linewidth]{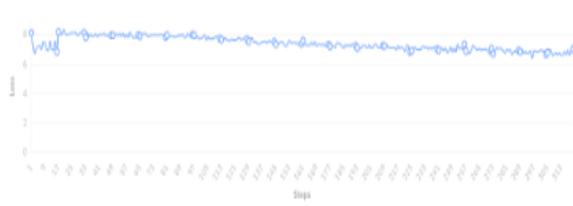}
    \caption{Prompt tuning results (size = 9950).}
    \label{fig:tuningresult-9950}
\end{figure}

As shown in Figure~\ref{fig:tuningresult-250}, when there is little training data, the loss decreases slowly, but when a sufficient amount of training data is used, the loss decreases quickly, as shown in Figure~\ref{fig:tuningresult-9950}. This shows that this is an effective optimization method in an environment where there is sufficient learning data in advance and prompt tuning such as Tuning Studio can be applied.

\subsection{Auto Prompt-Engineering}

We implemented a mechanism based on Automatic Prompt Engineer~\cite{Zhou+:ICLR-23:autoPE}  and confirmed that it is possible to generate prompts using LLM. However, in this verification, we were unable to generate prompts that were significantly better than those generated by humans. This may be because prompt optimization is required to generate prompts, and the optimization method depends on each model.

In order to utilize automatic prompt engineering in business, it is necessary to develop a more versatile and maintainable solution. In short, it would be sufficient to fix the LLM for generating prompts and perform meta-prompt engineering, but this is only applicable in limited cases. In a more general sense, a mechanism is needed that allows prompt tuning of meta-prompts. 

\subsection{Few-shot Learning}

Next, we verified few-shot learning, which allows for efficient tuning with a small amount of training data. The data injected into the prompts was the same as the dataset used for prompt tuning, and example sentences were randomly selected from that dataset, and the change in loss due to the number of injections was compared. The results are shown in Figure 4. The target models for comparison were ibm/granite-13b-chat-v2, as well as two models that had been subjected to prompt tuning.

For models that have not undergone prompt tuning, it has been confirmed that loss is reduced when few-shot learning is applied to zero-shot. However, when the number of injected samples exceeds a certain level, loss tends to increase, and it is necessary to set an optimal number of injected samples.

In contrast, the same model with prompt tuning showed lower loss than the model without prompt tuning at the zero-shot point, but when the sample was injected at the few-shot point, the loss tended to increase. This phenomenon will be discussed in the next chapter.

\subsection{Pitfalls}

As in the example of Few Shot Learning in the previous chapter, when prompt tuning and other tuning methods were applied to LLM, cases were confirmed in which the effects were negated depending on the combination and order. This is because additional optimization inputs prompts with different patterns from the learning data used in prompt tuning, which is thought to be a phenomenon similar to overfitting in machine learning.

To deal with such cases, it is necessary to carefully select the appropriate combination of tuning methods and the order in which they are applied. For example, in manual and automatic prompt engineering and few shot learning, it is expected that both tuning methods will be effective if prompt tuning is performed after determining the base prompt.

In addition, prompt tuning is not a very effective tuning method when combined with methods that dynamically change prompts, such as RAG (Retrieval-Augmented Generation), and it is considered preferable to use these methods exclusively.

\section{Concluding Remarks}

In this paper, we presented the problem of LLM changes in the development and operation of generative AI applications as generative AI model agility, and discussed solutions. We defined GenAIOps, a generative AI application development and operation process that extends MLOps, and clarified the issues and related matters within each step. We took up prompt tuning as a solution direction and investigated existing tools and research. In particular, we discussed the effectiveness and limitations from a case study that utilized existing soft prompt tuning tools.

There are many directions for future research, but from the perspective of generative AI model agility, we would like to clarify the compatibility between existing LLMs. We also want to develop practical tools for advanced prompt tuning, which will be essential for CI/CD of GenAIOps.
%\begin{small}
\bibliographystyle{plain}
\bibliography{bibs/mlops,bibs/autope,bibs/fairness}
%\end{small}

\end{document}